\documentclass[10pt,conference]{IEEEtran} 
\IEEEoverridecommandlockouts
\usepackage{url}

\usepackage{cite}
\usepackage{amsmath,amssymb,amsfonts}
\usepackage{algorithmic}
\usepackage{graphicx}
\usepackage{textcomp}
\usepackage{xcolor}
\usepackage{multirow}
\usepackage{longtable}
\usepackage{listings}
\usepackage[text=Pre-Print,colorspec=0.9]{draftwatermark}
\def\BibTeX{{\rm B\kern-.05em{\sc i\kern-.025em b}\kern-.08em
    T\kern-.1667em\lower.7ex\hbox{E}\kern-.125emX}}

\newcommand{\realsearchgoal}{The goal of this paper is to aid software engineering educators in designing a comprehensive software security course by sharing an experience running a software security course for the eleventh time.}
\begin{document}

\title{Structuring a Comprehensive Software Security Course Around the OWASP Application Security Verification Standard
\thanks{\copyright 2021 IEEE. Personal use of this material is permitted. Permission
from IEEE must be obtained for all other uses, in any current or future
media, including reprinting/republishing this material for advertising or
promotional purposes, creating new collective works, for resale or
redistribution to servers or lists, or reuse of any copyrighted
component of this work in other works. This material is based upon work supported by the National Science Foundation under Grant No. 1909516.  Any opinions, findings, and conclusions or recommendations expressed in this material are those of the author(s) and do not necessarily reflect the views of the National Science Foundation.} }

\author{\IEEEauthorblockN{Sarah E. Elder}
\IEEEauthorblockA{\textit{Department of Computer Science)} \\
\textit{North Carolina State University}\\
Raleigh, USA\\
seelder@ncsu.edu}
\and
\IEEEauthorblockN{Nusrat Zahan}
\IEEEauthorblockA{\textit{Department of Computer Science)} \\
\textit{North Carolina State University}\\
Raleigh, USA\\
nzahan@ncsu.edu}
\and
\IEEEauthorblockN{Val Kozarev}
\IEEEauthorblockA{\textit{Department of Computer Science} \\
\textit{North Carolina State University}}
\and
\IEEEauthorblockN{Rui Shu}
\IEEEauthorblockA{\textit{Department of Computer Science} \\
\textit{North Carolina State University}\\
Raleigh, USA\\
rshu@ncsu.edu}
\and
\IEEEauthorblockN{Tim Menzies}
\IEEEauthorblockA{\textit{Department of Computer Science} \\
\textit{North Carolina State University}\\
Raleigh, USA \\
timm@ieee.org}
\and
\IEEEauthorblockN{Laurie Williams}
\IEEEauthorblockA{\textit{Department of Computer Science} \\
\textit{North Carolina State University}\\
Raleigh, USA \\
laurie\_williams@ncsu.edu}
}

\maketitle



\begin{abstract}
  Lack  of  security  expertise  among  software  practitioners  is a  problem  with  many  implications.  First,  there  is  a  deficit of security professionals to meet current needs. Additionally, even  practitioners  who  do  not  plan  to  work  in  security  may benefit  from  increased  understanding  of  security. 
  \emph{\realsearchgoal}
 Through all the eleven years of running the software security course, the course objectives have been comprehensive -- ranging from security testing, to secure design and coding, to security requirements to security risk management.  For the first time in this eleventh year, a theme of the course assignments was to map vulnerability discovery to the security controls of the Open  Web  Application  Security  Project  (OWASP)  Application  Security  Verification  Standard (ASVS). Based upon student performance on a final exploratory penetration testing project, this mapping may have increased students' depth of understanding of a wider range of security topics.  The students efficiently detected 191 unique and verified vulnerabilities of 28 different Common Weakness Enumeration (CWE) types during a three-hour period in the OpenMRS project, an electronic health record application in active use.       
\end{abstract}
\section{Introduction}
Throughout the world, demand exceeds supply for trained cybersecurity professionals.  In the United States, the cybersecurity workforce  has been deemed, in a Presidential Executive Order \cite{TrumpEO}, a ``strategic asset that protects the American people, the homeland, and the American way of life.'' The Cyberseek\footnote{\url{https://www.cyberseek.org}} project, sponsored by the US National Institute of Standards and Technology (NIST), highlights a dangerous shortage of cybersecurity workers that puts the country's digital privacy and infrastructure at risk. To help address this problem, NIST has created the National Initiative for Cybersecurity Education (NICE) Cybersecurity Workforce Framework \cite{NICEOriginal}.  The mission of the NICE Framework is to ``energize and promote a robust network and an ecosystem of cybersecurity education, training, and workforce development.''  The NICE Framework establishes a taxonomy and common lexicon for describing the cybersecurity workforce including tasks, skills, training, and personnel; as well as a matrix of specialty areas for K-12 through university degree programs.  Similarly, the United Kingdom National Cyber Security Programme supported the development of the Cybersecurity Body of Knowledge (CyBoK)\footnote{\url{https://www.cybok.org/}}. The CyBoK codifies literature, such as textbooks, academic research articles, technical reports, white papers and standards on 19 essential knowledge areas of cybersecurity.

Higher education plays a key role in filling the gaps in the cybersecurity workforce.  In 2017, the two major computer science professional societies, the Association for Computing Machinery (ACM) and the IEEE Computer Society (IEEE-CS), published Curriculum Guidelines for Post-Secondary Degree Programs in Cybersecurity (CSEC2017 v1.0) \cite{Curricula}.
A key aspect of cybersecurity education is the area of \textit{software security} \cite{mcgraw} which is considered to be the intersection of software engineering and cybersecurity in which software engineers ``build security in'' to products.

\textit{\realsearchgoal} When this course was first offered in 2009, few courses on software security existed.  Through all the eleven years of running the software security course, the course objectives have been comprehensive – ranging from security testing, to secure design and coding,  to  security  requirements to security risk management. 

The course has always been designed to be experiential, whereby students applied classroom topics on a system. Due to the rapid evolutionary nature of cybersecurity, each year the course was updated to include new tools, new techniques, new standards, new regulations, and new research results.  In Spring 2020, a theme of the course assignments was to map vulnerability discovery to the security requirements of the Open  Web  Application  Security  Project  (OWASP) OWASP  Application  Security  Verification  Standard (ASVS).  \textit{Of all the changes made to the course offering over the eleven years, the change toward structuring the course around the ASVS seemed to be the most beneficial from a student learning perspective}, and we share our experiences in this paper.

With this paper, we make the following contributions:
\begin{itemize}
    \item A software security course structure (topics and assignments) that has matured over an eleven year period; and
    \item Our experience structuring a software security course around the OWASP Application Security Verification Standard (ASVS).
\end{itemize}

The rest of this paper is organized as follows.  In Section \ref{sec:relwork} and Section \ref{key}, we present related work and key concepts, respectively.  In Section \ref{sec:OWASP} we provide background on the ASVS standard which was the main innovative aspect added to the class. We lay out the course topics in Section \ref{Topics} and course assignments in Section \ref{sec:assignments}. We present the results of the final exploratory penetration testing exercise in Section \ref{sec:EPT Results}.  Finally, we present our lessons learned in Section \ref{sec:LessonsLearned}.

\section{Related Work}
\label{sec:relwork}
Previous work has been done on tools and techniques for security education. For example, work has been done in the use of serious games and the gamification of security research, particularly for the purpose of education including two summits on \emph{Gaming, Games, and Gamification in Security Education} (3GSE)\footnote{https://www.usenix.org/conferences/byname/885} and more recent works by Antonaci et al and {\v{S}}v{\'a}bensk{\`y}  et al \cite{antonaci2017gamification,vsvabensky2018enhancing}. Other work has looked at the use of Massive Open Online Courses (MOOCs) to teach cybersecurity\cite{antonaci2017gamification,theisen2017teaching,laato2020ai}. Many organizations provide online cybersecurity learning resources such as the Cyber Security Body of Knowledge (CYBOK)\footnote{\url{https://www.cybok.org/}} and SAFECode\footnote{\url{https://safecode.org/training/}}. Several researchers, teachers, and organizations provide lists of knowledge objectives or course outcomes such as work by Mead et al \cite{mead2010software} and the ACM/IEEE/AIS SIGSEC/IFIP Cybersecurity Curricula Guideline \cite{CybersecurityCurricula2017} which map knowledge objectives and learning outcomes to Bloom's Taxonomy.

Two studies \cite{ruef2016build,votipka2020understanding} from the University of Maryland (UMD) are particularly useful in understanding our own experiences in  Software Security education. The studies from UMD examine how students perform in a \textit{Build It, Break It, Fix It} competition where teams of students attempt to build the most secure software (the ``Build It'' phase), find the most vulnerabilities in each other's software (the ``Break It'' phase), and fix the vulnerabilities in their own software that have been found (the ``Fix It'' phase). The studies found that the reason most teams implemented security vulnerabilities was lack of knowledge. However, the impact of education on the vulnerabilities introduced by students was mixed. The first study, published in 2016 by Ruef et al.\cite{ruef2016build}, found that students who had taken an online course on cybersecurity performed better than students who had not taken the course. The second study found that while knowledge, or lack thereof, appeared to be the main reason vulnerabilities were introduced in the ``Build It'' phase of the competition, participation in the online course on cybersecurity did not seem to influence whether students introduced vulnerabilities. The authors suggest that experience with a variety of different types of vulnerabilities may help in  reducing the number and types of vulnerabilities introduced (in the ``Build it'' phase), as well as increasing the number and types of vulnerabilities identified when looking for vulnerabilities in software (in the ``Break it'' phase). The results of these studies indicate the need for more strategies for security education. Using a comprehensive standard, such as ASVS, may be a helpful strategy, since students who utilize the ASVS framework should be more familiar with a wide variety of vulnerability types.

Other studies have examined the role of knowledge in software verification and validation. Votipka et al's study on \textit{Hackers vs Testers}\cite{votipka2018hackers} examines how white-hat hackers approach security testing as compared to software testing experts. The authors performed a series of interviews with white-hat hackers and with testers. Based on the interviews, the authors developed a model for the vulnerability discovery process. They found that although both hackers and testers follow similar processes for finding vulnerabilities, different experience and knowledge between the two groups led to different results. In a non-security-specific setting, Itkonen et al\cite{itkonen2013role} found that domain, system, and general software engineering knowledge were all useful in performing exploratory testing; and the ability to apply that knowledge ``on the fly'' may contribute to exploratory testing being more efficient than systematic, scripted testing.

This paper adds to existing work by providing our experience and lessons learned when using a standard, such as ASVS, to structure assignments and lectures in a graduate-level software security course at a public research university. As can be seen in Section \ref{Topics}, ASVS was incorporated into existing course objectives and topics. The course website\footnote{https://sites.google.com/a/ncsu.edu/csc515-software-security/Schedule-of-Subjects/Prelim} references many additional materials that were also used to help students. 

\section{Key Concepts} \label{key}
In this section, we define terminology needed to understand the rest of this paper.
\subsection{Vulnerability} 
\label{sec:def-vulnerability}
In the course, students were introduced to several definitions of the term \emph{security vulnerability}. While we do not think that the differences between the definitions would impact the student results, we provide these definitions for the reader. Two informal definitions used at the beginning of the course were that a security vulnerability is ``a bug with security consequences'' or ``a design flaw or poor coding that may allow an attacker to exploit software for a malicious purpose''. More formally, students were given the definition of a vulnerability from the U.S. National Institute of Standards and Technology (NIST) Security and Privacy Controls for Federal Information Systems and Organizations Special Publication (SP) 800-53\cite{NIST800-53R4}. NIST SP 800-53 defines a vulnerability as ``a weakness in ... an implementation that could be exploited by a threat source''. 
The vulnerability counts in Section \ref{sec:EPT-results} abide by the definition of vulnerability from the Common Vulnerabilities and Exposures (CVE)\footnote{https://cve.mitre.org} list that is also used by the U.S. National vulnerability Database\footnote{https://nvd.nist.gov/vuln}. Namely, that a vulnerability is a ``flaw in a software, firmware, hardware, or service component resulting from a weakness that can be exploited, causing a negative impact to the confidentiality, integrity, or availability of an impacted component or components''\footnote{https://cve.mitre.org/about/terminology.html}.


\subsection{Common Weakness Enumeration (CWE)} 
\label{sec:def-cwe}
Per the CWE  website, ``CWE is a community-developed list of software and hardware weakness types.''\footnote{https://cwe.mitre.org/}. Many security tools, such as the OWASP Application Security Verification Standard (ASVS) and most vulnerability detection tools, use CWEs to identify the types of vulnerabilities relevant to a given security requirement, test case, or tool alert. Each CWE type header describes the vulnerability, and each type has a unique numeric value which we will refer to as the CWE number. We use the vulnerability types in the CWE list to better understand and describe the diversity of vulnerabilities students found during their exploratory testing exercise in section \ref{sec:EPT Results}

\section{OWASP Application Security Verification Standard (ASVS)}
\label{sec:OWASP}

The Open Web Application Security Project (OWASP)\footnote{\url{https://owasp.org/}} is a non-profit organzation that promotes software security through educational resources, open-source software, and other open-source projects. 

The OWASP Application Security Verification Standard (ASVS) is an open standard for performing Web application security verification. The ASVS provides a high-level set of ``requirements or tests that can be used by architects, developers, testers, security professionals, tool vendors, and consumers to define, build, test and verify secure applications''\cite{OWASP2019ASVS}. The ASVS Project\footnote{https://owasp.org/www-project-application-security-verification-standard/} intends for the standard to be ``commercially workable'', and to provide adequate coverage so that ASVS can be used for a wide range of purposes including as an internal metric for security, guidance for developers when implementing new security features, or as a security baseline when evaluating third-party software or development contracts. The current version of OWASP ASVS is Version 4.0.1 released in March 2019\footnote{https://github.com/OWASP/ASVS/tree/v4.0.1}.

ASVS requirements are phrased in terms of what would need to be verified. The requirements must be adapted to a specific system under test to be effective. The requirements are grouped into higher-level sections and sub-sections. Additionally, each ASVS requirement is mapped to a CWE type using the CWE number. Where applicable, ASVS requirements are also mapped to applicable standards from the U.S. National Institute of Standards and Technology (NIST). An example ASVS requirement, ASVS 2.2.1, is shown in Figure \ref{fig:ASVSreqExample}. ASVS 2.2.1 is part of ASVS section \emph{V2: Authentication Verification Requirements}, subsection \emph{V2.2 General Authenticator Requirements}. ASVS requirements from section V2 are mapped to \textit{NIST Special Publication (SP) 800-63: Digital Identity Guidelines}. A violation of ASVS 2.2.1 would result in a vulnerability of the type \textit{CWE-307: Improper Restriction of Excessive Authentication Attempts}. Additionally, ASVS 2.2.1 maps to sections 5.2.2, 5.1.1.2, 5.1.4.2, and 5.1.5.2 of NIST SP 800-63.

\begin{figure*}[htb]
\caption{ASVS requirement 2.2.1}\label{fig:ASVSreqExample}
\vspace*{5mm}
\centering
\renewcommand{\arraystretch}{1.3}
\begin{tabular}{ p{18pt} | p{330pt} | p{10pt} | p{10pt}| p{10pt} | p{22pt}| p{30pt}}
\# & Description & L1 & L2 & L3 & CWE & NIST\S\\\hline\hline
2.2.1 & Verify that anti-automation controls are effective at mitigating breached credential testing, brute force, and account lockout attacks. Such controls include blocking the most common breached passwords, soft lockouts, rate limiting, CAPTCHA, ever increasing delays between attempts, IP address restrictions, or risk-based restrictions such as location, first login on a device, recent attempts to unlock the account, or similar. Verify that no more than 100 failed attempts per hour is possible on a single account. & \checkmark & \checkmark & \checkmark & 307 & 5.2.2/ 5.1.1.2/ 5.1.4.2/ 5.1.5.2\\
\end{tabular}
\end{figure*}

The OWASP ASVS has three levels of requirements. If a requirement falls within a level, it also falls within higher levels. For example a requirement that is part of Level 2 is also part of Level 3 of the standard. ASVS describes Level 1 as ``the bare minimum that any application should strive for''\cite{OWASP2019ASVS}. As of Version 4 of ASVS, Level 1 requirements can all be verified manually. Level 2 should be addressed by any applications handling sensitive information and is appropriate for most applications. Addressing Level 3 requirements may require more resources than organizations are willing or able to spend on some systems. However, Level 3 requirements should be met for critical systems. In the example in Figure \ref{fig:ASVSreqExample}, ASVS 2.2.1 is an L1 requirement, which means it must also be verified to  meet L2 and L3 requirements.


\section{Course Topics} \label{Topics}
The graduate-level Software Security course introduces students to the discipline of designing, developing, and testing secure and dependable software-based systems. The course website and resources can be accessed at \url{https://sites.google.com/a/ncsu.edu/csc515-software-security/} 

The four course learning objectives and topics taught to achieve these objectives are as follows:
\begin{enumerate}
    
\item \textbf{Security testing.} \textit{Objective 1:  Students will be able to perform all types of security testing.}  
\begin{itemize}
    \item \textbf {Penetration Testing (PT)}:  Penetration testing is a method for ``gaining assurance in the security of an IT system by attempting to breach some or all of that system's security, using the same tools and techniques as an adversary might.''\cite{PenetrationTestingDefinition}  In the class, students performed \underline{exploratory} penetration testing in which the tester ``spontaneously designs and executes tests based on the tester's existing relevant knowledge''\cite{2013ISO29119-1}. The students also performed \underline{systematic} penetration testing in which the tester methodically develops and documents  a test plan in which the test cases comprehensively address the security objectives of the system. The test plan is then manually executed against the System Under Test (SUT)\cite{smith2011systematizing,austin2011onetechniquenotenough,smith2012effective,austin2013comparison} in academic literature and in practice, penetration testing may refer to any dynamic security testing, automated or manual\cite{cruzes2017security,scandariato2013static,hafiz2016game}.  However, in practice  testers may use a combination of exploratory and systematic testing  \cite{votipka2018hackers}. Students also learned how to augment their testing with the use of a proxy server, OWASP ZAP\footnote{https://www.zaproxy.org/}, that allows the user to manipulate traffic outside the browser.
      Penetration testing can also be referred to as \textit{pentesting} and \textit{ethical hacking}.
    
    \item \textbf{Dynamic Application Security Testing (DAST)}: DAST tools automatically generate and run a test cases against the SUT, without any access to the SUT source code\footnote{\url{https://insights.sei.cmu.edu/sei_blog/2018/07/10-types-of-application-security-testing-tools-when-and-how-to-use-them.html}}.  DAST can also be be referred to as \textit{fuzzing}.
    \item \textbf{Static Application Security Testing (SAST)}: SAST tools automatically scan application source code for defects\cite{cruzes2017security,hafiz2016game,scandariato2013static}.  SAST can also be referred to as \textit{static analysis}.
    \item \textbf{Interactive Application Security Testing (IAST)}. IAST tools perform dynamic analysis but also have access to source code. IAST tools require a tester to interact with the application while the IAST tool monitors what code is exercised during the interactions\footnote{https://www.synopsys.com/glossary/what-is-iast.html}.
\end{itemize} 
\item \textbf{Secure design and coding practices.} \textit{Objective 2:  Students will understand secure design and coding practices to prevent common vulnerabilities from being injected into software.} 
\begin{itemize}
    \item \textbf{Design}. Approximately half of the vulnerabilities in a software product are due to design flaws \cite{mcgraw} that can cause architectural-level security problems and need to be fixed via redesigning a portion of a product.  The students learned about avoiding the ``Top 10 Software Security Design Flaws'' \cite{IEEECenter} according to the IEEE Center for Secure Design.
    \item \textbf{Coding}.  Approximately half of the vulnerabilities in a software system are due to implementation bugs that are caused by code-level security problems. Implementation bugs are usually fixed via altering some lines of code.  The students learned about avoiding common implementation bugs such as OWASP Top 10 Web Application Security Risks\footnote{https://owasp.org/www-project-top-ten/} and CWE Top 25 Most Dangerous Security Weaknesses\footnote{https://cwe.mitre.org/top25/archive/2020/2020\_cwe\_top25.html}.   
    \item \textbf{Process}. Students learned about organizing their secure design and coding practices around a Secure Development Lifecycle such as the Microsoft Secure Development Lifecycle (SDLC)\footnote{https://www.microsoft.com/en-us/securityengineering/sdl}.  Students also learned about assessing the SDLC and/or the security practices used by an organization via the Building Security In Maturity Model (BSIMM)\footnote{https://www.bsimm.com/} and the OWASP Software Assurance Maturity Model (SAMM)\footnote{https://www.opensamm.org/}.
\end{itemize}
\item \textbf{Security requirements}. \textit{Objective 3:  Students will be able to write security and privacy requirements.}  
Security and privacy requirements include compliance with necessary standards and regulations (such as GDPR\footnote{\url{https://gdpr-info.eu/}} or HIPAA\footnote{\url{https://www.hhs.gov/hipaa/index.html}}).  These security and privacy requirements must proactively include functionality to thwart attacker's attempts to exploit system functionality.  In the class, students learned about formal security requirements, such as those that are specified in NIST 800-53 Security and Privacy Controls\footnote{https://csrc.nist.gov/publications/detail/sp/800-53/rev-5/final}. Additionally, students used the following four techniques to aid in their ability to ``think like an attacker'' and to specify this defensive security functionality.  

\begin{itemize}
    \item \textbf{Adversarial thinking.} Students were exposed to the tactics, techniques of procedures (TTP) of attackers through the use of the MITRE ATT\&CK®\footnote{https://attack.mitre.org/}, a globally-accessible knowledge base of real-world observations and analysis of the actions of adversaries.
    \item \textbf{Threat modeling.} Through the use of threat modelling \cite{ThreatModel}, students considered, documented and discussed the security implications of design in the context of their planned operational environment and in a structured fashion. Threats could be enumerated using a systematic approach of considering each system component relative to the STRIDE (Spoofing, Tampering, Repudiation, Information Disclosure, Denial of Service, Elevation of Privilege) Model \cite{stride}. Threat models could also be collaboratively developed via the Elevation of Privilege game\footnote{\url{https://www.usenix.org/conference/3gse14/summit-program/presentation/shostack}}.  
    \item \textbf{Attack trees.} Students created conceptual diagrams of threats on systems and possible attacks to reach those threats \cite{attacktrees}.
    \item \textbf{Abuse cases.} Students created abuse cases \cite{Abuse}, which describe the system’s behavior when under attack by a malicious actor. To develop abuse cases, students enumerated the types of malicious actors who would be motivated to attack the system. 
\end{itemize}

\item \textbf{Security risk management.} \textit{Objective 4:  Students will be able to assess the security risk of a system under development.} Students learned to assess security risk in and a standardized and an informal way.
\begin{itemize}
    \item \textbf{Formal risk assessment.} Students learned to assess security risk using processes outlined in the NIST Cybersecurity Framework\footnote{https://www.nist.gov/cyberframework} and NIST 800-30 Guide for Conducting Risk Assessments\footnote{https://www.nist.gov/privacy-framework/nist-sp-800-30}. 
    \item \textbf{Informal, collaborative risk assessment.} Students played Protection Poker \cite{ProtectionPoker1, ProtectionPoker2}, an informal game for threat modeling, misuse case development, and relative security risk quantification. 
\end{itemize}
\end{enumerate}

\section{Course Assignments}
\label{sec:assignments}

In this section, we provide information on the four-part project the students completed.  We first provide information on the system the students worked on, followed by the structure of the assignments.

\subsection{System Under Test (SUT)} \label{OpenMRS}

The SUT is OpenMRS, an open-source medical records system designed to be flexible so it could be adapted to many contexts where other medical records systems were not available\footnote{https://atlas.openmrs.org/}. OpenMRS has over 900,000 lines of code.  OpenMRS, as described by the developer's manual is a ``Java-based web application capable of running on laptops in small clinics or large servers for nation-wide use''\footnote{\url{http://devmanual.openmrs.org/en/Kick_off/solving_the_health_it_challenges_our_responsemd.html}}. OpenMRS uses both java and javascript. The OpenMRS architecture is modular. 

\subsection{Equipment}
\label{sec:method-sut-equipment}
All student tasks were performed on a virtual machine using the school's virtual computing lab\footnote{http://vcl.apache.org/}. The use of a virtual computing lab allowed researchers to create a system image including the SUT (OpenMRS) and necessary testing tools. Each student could then checkout their own copy of the SUT and tools with minimal installation effort on the part of the student. For client-server tools, the server was setup in a separate VCL instance in advance by members of the research team with assistance from the teaching staff. All students accessed the same server instance through different accounts. Student images were assigned 4 cores, 8G RAM, and 40G disk space. One server instance had 4 cores, 8G RAM, and 60G disk space. The other server instance had 8 cores, 16G RAM, and 80G disk space. These specifications were based on the minimum requirements needed for students to complete the tasks in their assignments. Full testing of the entire OpenMRS system was outside the scope of student assignments.


\subsection{Assignment Structure}

Through the semester, the students completed a four-part project in teams of 3-4 students.   The four parts are as follows.
\subsubsection{Project Part 1}
This assignment had two sub-parts, both involving the ASVS.
\begin{itemize}
\item \textbf{Systematic Penetration Testing (SPT).}  The students planned and executed 15 penetration test cases.  Each test case had to map to a unique ASVS Level 1 or Level 2 control.  Student reported whether the test cases passed or failed, and the CWE of the vulnerability that was being tested for.  Most teams had failing test cases, indicating they found vulnerabilities in OpenMRS.  
\item \textbf{SAST.}  Students ran two commercially-available SAST tools on OpenMRS:  Fortify\footnote{\url{https://www.microfocus.com/en-us/products/static-code-analysis-sast/overview}} and Synposys Coverity\footnote{\url{https://www.synopsys.com/software-integrity/security-testing/static-analysis-sast.html}}.  For each tool, students randomly chose 10 alerts produced by the tool and determined if the alert was a true or false positive.  If the alert was a true positive, the student had to explain how to fix the vulnerability, map the vulnerability to the CWE, and map the vulnerability to an ASVS control. 
\end{itemize}

\subsubsection{Project Part 2}
This assignment had two sub-parts, the first of which involved the ASVS.
\begin{itemize}
    \item \textbf{DAST.} The students ran the OWASP ZAP DAST and a commercially-available DAST, Synopsys Defensics\footnote{\url{https://www.synopsys.com/software-integrity/security-testing/fuzz-testing.html}}. The students chose five true positive alerts produced by each tool.  For each of these, they wrote a structured black box text to replicate the discovered vulnerability.  Each of these test cases were mapped to the associated ASVS control and CWE.
    \item \textbf{Vulnerable Dependencies.} Modern software uses many third-party libraries and frameworks as dependencies. Known vulnerabilities in dependencies is a major security risk. Students ran five tools (OWASP-Dependency-Check\footnote{\url{https://owasp.org/www-project-dependency-check/}}, RedHat Victims\footnote{\url{https://github.com/victims/maven-security-versions}}, GitHub Security scanners\footnote{\url{https://github.com/features/security}}, Sonatype DepShield\footnote{\url{https://depshield.github.io/}}, and Snyk\footnote{{https://snyk.io/}}) to identify the vulnerable dependencies in OpenMRS.  Students compared the output of the five tools.
\end{itemize}
\subsubsection{Project Part 3}
This assignment had four sub-parts, the first three of which involved the ASVS directly.
\begin{itemize}
    \item \textbf{Logging.} Students wrote 10 black box test cases for ASVS Section \emph{V7: Error Handling and Logging Verification Requirements} requirements for Levels 1 and 2.
    \item \textbf{IAST.}  Students ran the Synopsys Seeker tool \footnote{https://www.synopsys.com/software-integrity/security-testing/interactive-application-security-testing/demo.html} using five failing black box test cases from their earlier work to seed the Seeker run.  For five of the true positive vulnerabilities identified by the tool, the students had to write a black box test case to replicate each discovered vulnerability. Each of these black box test cases were mapped to the associated ASVS control and CWE. 
    \item \textbf{Test coverage.}  ASVS has 14 sections.  Students computed their test coverage for each of these sections for all the test cases they had written during Parts 1-3 of the course.  The students then wrote 5 more test cases to increase the coverage of ASVS controls that they did not have a test case for.  
    \item \textbf{Vulnerability discovery comparison.}  Students compared the effectiveness and efficiency of the four detection techniques they had used (systematic penetration testing, SAST, DAST, and IAST).  They computed an efficiency measure based upon true positive vulnerabilities detected per hour. They also recorded all the CWEs detected by each tool.  Students were asked to reflect upon their experience with these techniques, comparing their ability of each technique to efficiently and effectively detect a wide range of types of exploitable vulnerabilities.
\end{itemize}
\subsubsection{Project Part 4}
This assignment had three sub-parts, the first two of which involved the ASVS directly.
\begin{itemize}
    \item \textbf{Protection Poker.} The students wrote 5 new functional requirements for OpenMRS to add functionality that is not in the system yet.  The played Protection Poker, using the OpenMRS database tables on these requirements and reflected how to reduce the security risk.
    \item \textbf{Vulnerability fix.} Students submitted a fix for a vulnerability that had been detected earlier in the class.
    \item \textbf{Exploratory Penetration Testing (EPT).} Students were assigned to individually spend three hours performing exploratory testing.   Students produced a video recorded of their three-hour session, noted any vulnerabilities found, and created black box test cases (labeled with ASVS control) based on the vulnerabilities found to document their results in a replicable way.
\end{itemize}


\section{Exploratory Penetration Testing Exercise:  A View into Student Learning}\label{sec:EPT Results}

While not a formal experiment and without a baseline comparison to a prior semester, we consider the student's performance on the final exploratory penetration testing exercise to be indicative of the skills they had learned during the semester.  We share the verbatim assignment and the student results in this section.

\subsection{Verbatim Assignment}
The text below is the exact text students were given on how to perform the final exploratory penetration testing exercise.\\

\textit{Each team member is to perform 3 hours of exploratory penetration testing on OpenMRS.  This testing is to be done opportunistically, based upon your general knowledge of OpenMRS but without a test plan, as is done by professional penetration testers.  DO NOT USE YOUR OLD BLACK BOX TESTS FROM PRIOR MODULES. Use a screen video/voice screen recorder to record your penetration testing actions.  Speak aloud as you work to describe your actions, such as, ``I see the input field for logging in.  I'm going to see if 1=1 works for a password.'' or ``I see a parameter in the URL, I'm going to see what happens if I change the URL.'' You should be speaking around once/minute to narrate what you are attempting.  You don't have to do all 3 hours in one session, but you should have 3 hours of annotated video to document your penetration testing.  There's lots of screen recorders available -- if you know of a free one and can suggest it to your classmates, please post on Piazza.} 

\textit{Pause the recording every time you have a true positive vulnerability.  Note how long you have been working so a log of your work and the time between vulnerability discovery is created (For example, Vulnerability \#1 was found at 1 hour and 12 minutes, Vulnerability \#2 was found at  1 hour and 30 minutes, etc.) If you work in multiple sessions, the elapsed time will pick up where you left off the prior session -- like if you do one session for 1 hour 15 minutes, the second session begins at 1 hour 16 minutes.  Take a screen shot and number each true positive vulnerability.  Record your actions such that this vulnerability could be replicated by someone else via a black box test case.  Record the CWE for your true positive vulnerability. Record your work as in the following table. The reference info for video traceability is to aid a reviewer in watching you find the vulnerability.  If you have one video, the ``time'' should aid in finding the appropriate part of the video.  If you have multiple videos, please specify which video and what time on that video. }

While students were graded on their video and results, only the results were studied further. We discuss the results in the next section.

\subsection{Data Collection} \label{sec:EPT-data}
The exploratory testing results were collected, alongside other data as part of a separate study on vulnerability detection techniques. Hence the data we discuss focuses on unique vulnerabilities rather than student responses. For the same reason, we did not retain the videos students created as they were not necessary and not easily anonymized. Student data was collected following North Carolina State University's Institutional Review Board Protocol 20569. The protocol was amended to verify that the EPT vulnerabilities could be discussed in this study. Sixty-three of seventy students  allowed their data to be used for the study by signing an informed consent form. The data was collected during the Spring Semester of 2020, the first semester the course was structured around ASVS. As part of the original vulnerability detection technique comparison, three researchers who are authors on this paper reviewed the student results remove erroneous vulnerability reports and duplicate vulnerabilities. 

\subsection{Students' Previous Experience}\label{sec:EPT-experience}
At the beginning of the course, students were asked to fill out a survey about their experience relevant to the course. The four survey questions were as follows:
\begin{itemize}
\setlength{\itemindent}{-0.1in}
    \item Q1: How much time have you spent working at a professional software organization -- including internships -- in terms of the \# of years and the \# of months?
    \item Q2: On a scale from 1 (none) to 5 (fully), how much of the time has your work at a professional software organization involved cybersecurity?
    \item Q3: Which of the follow classes have you already completed?  
    \item Q4: Which of the following classes are you currently taking? 
\end{itemize}

Q1 was short answer. For Q2, students selected a single number between 1 and 5. For Q3, the students could select as many options as were appropriate from a list of five graduate-level security or privacy courses and one undergraduate-level security course offered at the institution. For Q4, the students selected from the two graduate-level security courses that were also offered in the Spring of 2020. These two courses had both been offered previously, and were also part of the list for Q3.

Fifty-nine of the sixty-three students who agreed to let their data be used for the study had responded to the survey. Four students' responses to Q1 provided a numeric value but did not specify whether the numeric value indicated years or months, rendering the value unusable. For example, if a student simply put ``3'', we did not know if the student had 3 years of industry experience or 3 months of industry experience. Consequently, industry experience from 55 participants was used to approximate the average industry experience of the student participants.  

\begin{figure}[!t]
\centering \includegraphics[width=2.95in]{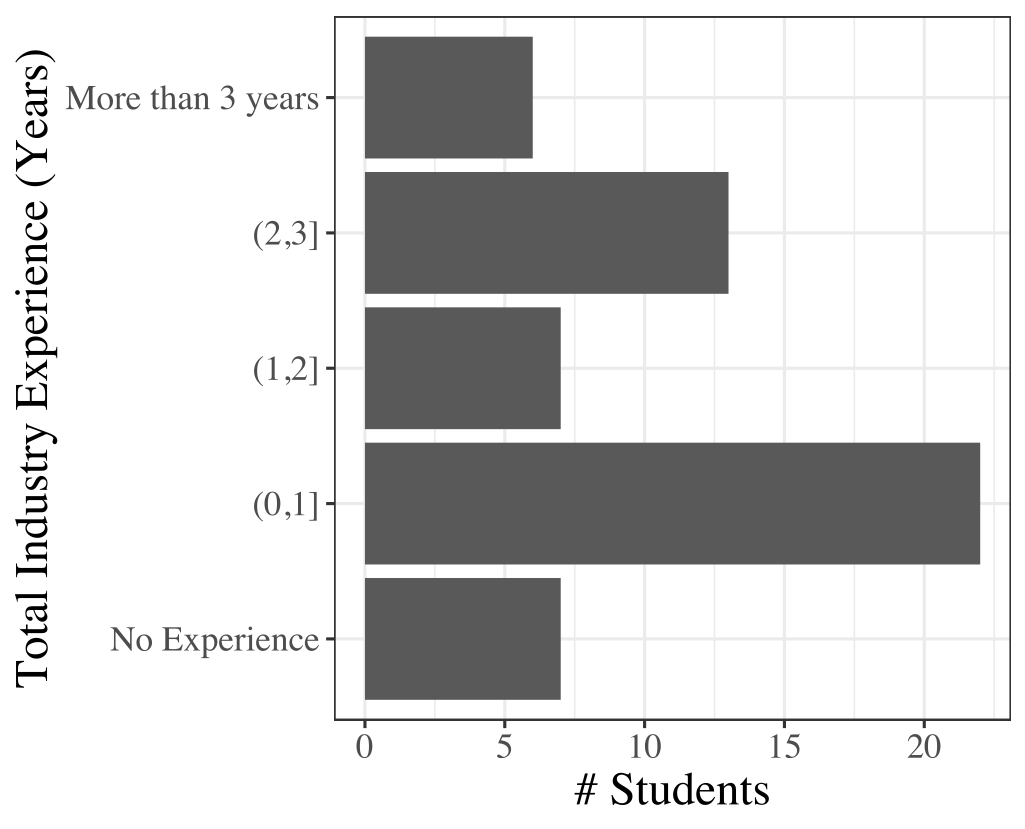} \caption[Student Industry Experience (Q1)]{Student Industry Experience (Q1)}
\label{fig:StudentExperience}
\end{figure}

Figure \ref{fig:StudentExperience} shows the student's industry experience (Q1) in years. We use set notation where $($ indicates exclusive and $]$ indicates inclusive. For example, in Figure \ref{fig:StudentExperience}, a student with exactly 2 years of experience would fall within the category $(1,2]$ and not within the category $(2,3]$. The median industry experience as indicated by answers to Q1 was 1 year. The average industry experience was 1 year 8 months. Of the 55 respondents whose answer for industry experience was clear, 7 had no industry experience at all.

\begin{figure}[!t]
\centering \includegraphics[width=2.95in]{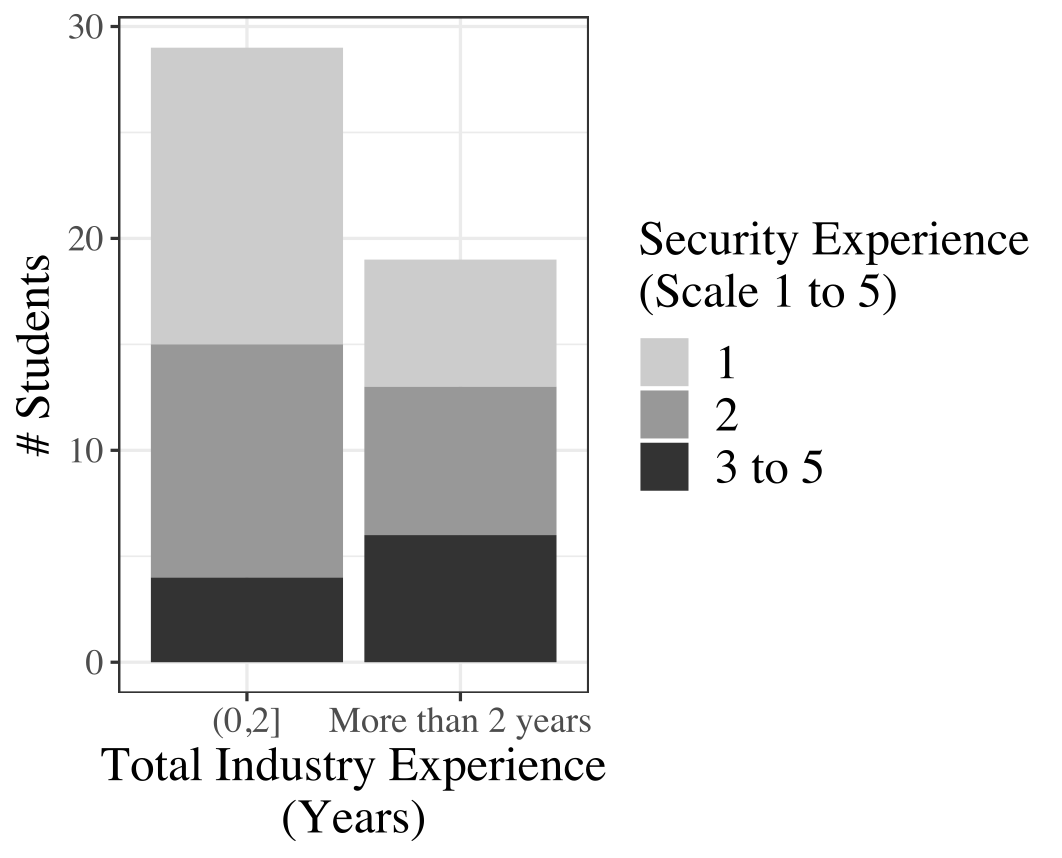} \caption[Security-Related Industry Experience (Q2)]{Security-Related Industry Experience (Q2)}
\label{fig:SecurityExperience}
\end{figure}

Figure \ref{fig:SecurityExperience} shows that, among the 48 students who had industry experience, most students had some exposure to security. In other words, most students answered at least 2 to Q2. However, 20 of the 48 students who had any industry experience had no industry experience relevant to security. Only 10 students indicated a 3 or higher in answer to Q2. Students with more industry experience also had more security experience. Possible reasons for this include students having more time to gain security-related experience when they have more industry experience generally, and students with more industry experience in security being more likely to select a security-related course than students who intend to follow a career elsewhere within computer science.

Additionally, of the 59 students who responded to the survey, only 8 had previously taken a course in security or privacy (Q3). Nine students were currently taking a course in security or privacy in addition to the course from which the data was collected (Q4).

\subsection{Student Results}\label{sec:EPT-results}
\begin{figure*}[htb]
\centering \includegraphics[width=7in]{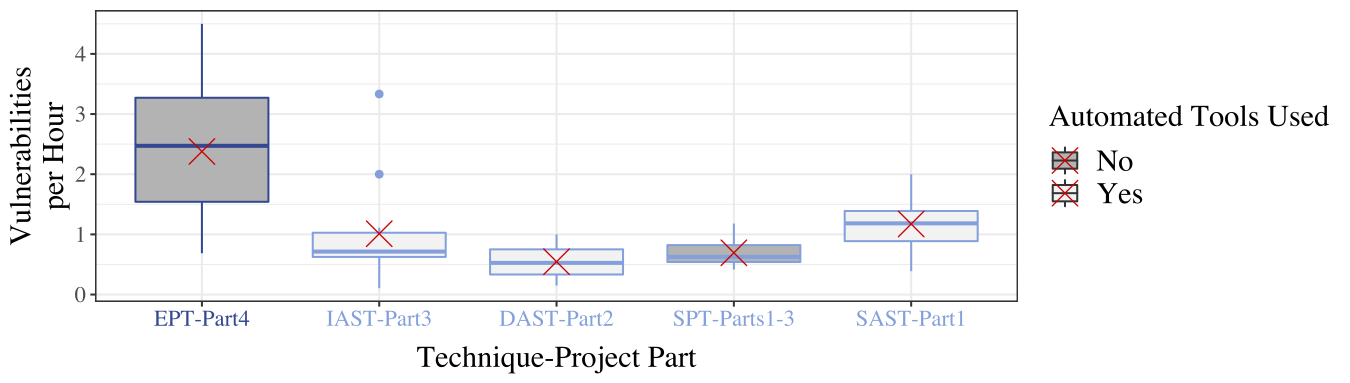} \caption[Exploratory Penetration Testing Efficiency]{Exploratory Penetration Testing Efficiency\footnotemark}
\label{fig:EPT_Efficiency}
\end{figure*}

Figure \ref{fig:EPT_Efficiency} shows student efficiencies as a boxplot. The data shown in Figure \ref{fig:EPT_Efficiency} has been trimmed\cite{wilcox2003modern,kitchenham2017robust}, using the median absolute deviation and median (MADN) to identify the most extreme outlier for each technique. The most extreme outlier for each technique was then removed from the dataset. Additionally, we use the abbreviations EPT and SPT to indicate exploratory penetration testing and systematic penetration testing respectively. 

As shown in Figure \ref{fig:EPT_Efficiency}, for the last assignment students were relatively efficient in using exploratory penetration testing to find vulnerabilities. Exploratory testing is highlighted in dark blue. The median, 2.47 vulnerabilities per hour in the case of exploratory testing, is indicated by the line in the middle of the boxplot. The average efficiency, 2.38 vulnerabilities per hour in the case of exploratory testing, is shown by the red x. Many factors clearly influenced student performance. Course information alone does not account for the difference since the vulnerability detection techniques used in Part 2 and Part 3 of the student project were less efficient than vulnerability detection techniques used in Part 1 as well as Part 4. However, the fact that the final exploratory testing assignment was the most efficient vulnerability detection assignment for the students is noteworthy.  As discussed in Section \ref{sec:relwork}, knowledge is thought to be a key factor in exploratory testing. Furthermore, as described in Section \ref{sec:EPT-experience}, a majority of students had little or no security experience prior to this course. Nevertheless, the students were more efficient with exploratory testing than with tool-based techniques (SAST, DAST, and IAST), or with systematic penetration testing. The relatively high efficiency of the students when applying exploratory testing at the end of the course is promising.

In addition to high efficiency, students were able to find many different types of vulnerabilities during their exploratory testing exercise. Table \ref{tab:UniqueVuln} shows the unique vulnerabilities found by students using exploratory testing alone. Students recorded over 450 vulnerabilities. Since teams worked independently, many teams coincidentally found the same vulnerability. Additionally some reported vulnerabilities were not actually vulnerabilities, e.g. a student reported an information disclosure concern over the database URL being accessible. However, this access was by an admin, on a page designed to give admin users access to this sensitive information.  In spite of the false positives and overlap, the student teams collectively found 191 unique vulnerabilities from 28 different CWE types. These 191 unique vulnerabilities are quantified in Table \ref{tab:UniqueVuln}.

\begin{table}
\renewcommand{\arraystretch}{1.3} \caption{Vulnerabilities Found through Exploratory Penetration Testing} \label{tab:UniqueVuln}
\vspace*{5mm}
\centering
\begin{tabular}{ p{200pt}  |  p{30pt} }
CWE & \# Unique Vulns\\\hline\hline
16 - Configuration & 2\\\hline
20 - Improper Input Validation & 13\\\hline
79 - Improper Neutralization of Input During Web Page Generation ('Cross-site Scripting') & 80\\\hline
200 - Information Exposure & 4\\\hline
209 - Information Exposure Through an Error Message & 10\\\hline
269 - Improper Privilege Management & 3\\\hline
272 - Least Privilege Violation & 1\\\hline
280 - Improper Handling of Insufficient Permissions or Privileges  & 1\\\hline
285 - Improper Authorization & 23\\\hline
308 - Use of Single-factor Authentication & 1\\\hline
319 - Cleartext Transmission of Sensitive Information & 4\\\hline
419 - Unprotected Primary Channel & 3\\\hline
434 - Unrestricted Upload of File with Dangerous Type & 1\\\hline
472 - External Control of Assumed-Immutable Web Parameter & 1\\\hline
509 - Replicating Malicious Code (Virus or Worm) & 1\\\hline
521 - Weak Password Requirements & 7\\\hline
532 - Information Exposure Through Log Files & 3\\\hline
544 - Missing Standardized Error Handling Mechanism & 5\\\hline
550 - Information Exposure Through Server Error Message & 1\\\hline
598 - Information Exposure Through Query Strings in GET Request & 6\\\hline
601 - URL Redirection to Untrusted Site ('Open Redirect') & 2\\\hline
602 - Client-Side Enforcement of Server-Side Security & 1\\\hline
613 - Insufficient Session Expiration & 3\\\hline
614 - Sensitive Cookie in HTTPS Session Without 'Secure' Attribute & 1\\\hline
620 - Unverified Password Change & 1\\\hline
639 - Authorization Bypass Through User-Controlled Key & 5\\\hline
770 - Allocation of Resources Without Limits or Throttling & 2\\\hline
778 - Insufficient Logging & 11\\\hline
\hline
Total & 191\\\hline
\end{tabular}
\end{table}

\section{Lessons Learned about the Use of ASVS} \label{sec:LessonsLearned}
 
\subsection{Improved Student Performance}
Although we do not have a control group to formally quantify an improved performance on efficiency and effectiveness in finding vulnerabilities, we were not expecting students to find as many unique vulnerabilities using exploratory penetration testing as they did, particularly in such a short timeframe. Student familiarity with the ASVS framework may have contributed their productivity.

\subsection{Assisted in Identifying Knowledge Gaps}
One advantage of ASVS when students erroneously identified or classified vulnerabilities was that the student references to ASVS could facilitate identifying knowledge gaps.  For example, several ASVS requirements are about the entropy needed to secure authentication information and other secrets, such as ASVS 3.2.2 which states ``Verify that session tokens possess at least 64 bits of entropy.''. Many students wrote systematic test cases or logged exploratory tests vulnerabilities against this requirement. Although most of the tests and findings correctly identified the session token, most of the tests or findings incorrectly measured 64 bits of entropy. By referencing the original ASVS, we were sometimes able to better understand the student's intent. Entropy is outside the scope of the course, and would be anticipated as a gap in this case. Other gaps, such as a gap in what would be covered by a prerequisite course, might be of greater interest.

\subsection{Motivated Self-Learning or Knowledge Sharing}
In the projects Parts 1-3, students wrote each test case based on a single ASVS control, typically writing one test case per control. The course did not provide the technical knowledge for many of the ASVS controls. Additionally, the ASVS assumes that the reader either already has some knowledge of security concepts, or will obtain that knowledge elsewhere. As a result, students may have had to do some research in order to write a test case for a control, such as on entropy to continue the above example, or to share knowledge about security technology among team members. 

\subsection{Continued Need to Emphasize Repeatable Test Cases}
In each project part, students were instructed to document their security test cases via ``Detailed and repeatable (the same steps could be done by anyone who reads the instructions) instructions for how to execute the test case''. This type of repeatable steps to replicate a vulnerability is necessary when reporting a discovered vulnerability to a product team. Since a Software Engineering course was a prerequisite to Software Security course, we assumed that students would be familiar with writing repeatable test cases. However, we found that students often did not specify their test cases to that level of detail. Since ASVS is high-level, to our knowledge ASVS alone cannot address this issue.

\subsection{Awareness of Security Resources}
Courses may instruct students on cybersecurity practices and technologies. However, government and practitioner organizations are creating valuable resources, such as the ASVS, and others mentioned in Section \ref{Topics}.  Students should learn to look for resources which can aid them in conducting their security work more systematically and based upon the input of security experts.

\section{Acknowledgments}
We thank Jiaming Jiang for her support as teaching assistant for this class.  We are grateful to the I/T staff at the university for their assistance in ensuring that we had sufficient computing power running course. We also thank the students in the software security class.  Finally, we thank all the members of the Realsearch research group for their valuable feedback through this project.

This material is based upon work supported by the National Science Foundation under Grant No. 1909516.  Any opinions, findings, and conclusions or recommendations expressed in this material are those of the author(s) and do not necessarily reflect the views of the National Science Foundation. 

\bibliographystyle{IEEEtran}
\bibliography{IEEEabrv,elder}
\end{document}